\documentstyle[preprint,aps]{revtex}
\begin{document}

\draft
\title
{\bf Light emitting single electron transistors}

\author{$^{1}$David M.-T. Kuo and $^{2}$ Yia-chung Chang}
\address{$^{1}$Department of Electrical Engineering, National Central
University, Chung-Li, Taiwan, 320, Republic of China}
\address
{$^2$Department of Physics\\
University of Illinois at Urbana-Champaign, Urbana, Illinois
61801}

\date{\today}
\maketitle

\begin{abstract}
The dynamic properties of light-emitting single-electron
transistors (LESETs) made from quantum dots are theoretically
studied by using nonequilibrium Green's function method. Holes
residing at QD created by small ac signals added in the base
electrode of valence band lead to the exciton assisted tunnelling
level for the electron tunnelling from emitter to collector, it is
therefore such small signals can be amplified. LESETs can be
employed as efficient single-photon detectors.
\end{abstract}

\newpage
\section{Introduction}
The single electron tunnelling devices made from quantum dots have
been extensively studied not only for interesting physics such as
Kondo effect$^{1}$ and Coulomb Blockade$^{2}$, but also for their
promising applications in single-electron transistors (SETs)$^{3}$
and single-photon generators (SPGs).$^{4}$ SPGs provide the
antibunching single-photon sources (SPSs) for being applied in
quantum communications. Not only do SPSs crucial, but also
single-photon detectors in the implementation of quantum
communication systems. Blakesley et al have demonstrated resonant
diodes as efficient single-photon detectors$^{5}$. Nevertheless,
tunnelling current through the quantum wells are not sensitive
enough to distinguish the intensity of SPS in the ns period.$^{5}$
To detect single-photon train in a short period, we proposal to
employ SETs to detect such a single-photon train.

The studied SET is shown in Fig. 1, where a single quantum dot
(InAs) embedded in the GaAs matrix connected with three terminals
(emitter, collector and base).We consider only the ground states
of conduction and valence band of the QD, which are above the
Fermi energy levels of three electrodes $E_{F,i=e,c,b}$.
Conventional SETs are consisted of source, drain and gate
electrodes, where the gate electrode is used to tune the energy
levels of QDs, but not supplies carriers. In contrast, the base
terminal will provide holes into the QD in this study. In the
absence of holes, the applied voltage crossing the emitter and
collector is insufficient to yield significant current. Once holes
residing at the QD creating new channels for electrons tunnelling
from emitter to collector, the remarkable collector current
($I_c$)is yielded. In addition, the base current ($I_b$) arises
from the photon emission via electron-hole recombination of the
exciton complex state will be observed.

Unlike traditional transistors (electric output) and light
emitting diodes (optical output), the light-emitting
single-electron transistors (LESETs) with electric and optical
outputs can readily reach high current gain ($\beta=I_c/I_b \gg
1$) (only voltage gain in conventional SETs). It is expected that
$\beta \gg1$ is a manifested demonstration of small signal
amplifier. Therefore, it could be applied to detect single-photon
train. For a small semiconductor QD, particle collection
significantly influences the transport and optical properties of
single electron tunnelling devices. Within the framework of the
effective mass model$^{6}$, Fig. 2 show  the electron-electron
interaction $U_{e}$, hole-hole interaction $U_{h}$  and
electron-hole interaction $U_{eh}$ for different exciton complex
configurations. The electron-hole Coulomb energy $U_{eh}$ in the
exciton $X$ is almost the same as that in $X^2$ and is omitted in
this plot.  Note that the magnitude of charging energies ($U_e$
and $U_h$) for electrons or holes in the same order of thermal
energy of room temperature $k_B T$, where $k_B$ is a Boltzmann
constant. This indicates that the operation temperature of system
should be much lower than room temperature.

Owing to the applied bias crossing the QD, the electric field
effect is not negligible in the variation of particle
interactions. We adopt the size of QD with radius 7.5 nm and
height 3 nm to study the electric field effect on the particle
Coulomb interactions. We assume that the z axis is directed from
the base to the apex of the dot. Fig. 3 shows the Coulomb
interaction strengths as functions of electric field for different
exciton complexes. We see that $U_{h}$, $U_{eh}$ and $U_e$ display
asymmetric behavior of electric field as a result of geometer of
the dots. Increasing electric field, the deduction of $U_{eh}$
indicates that the electron-hole separation increases. However, we
note that $U_h$ increases in the positive direction of electric
field since the wave functions of holes become more localize. Even
though the enhancement of $U_e$ is observed in the negative
direction of electric field, it only exists at very small electric
field region. When the electric field is larger than a threshold
value, the wave functions of electrons become delocalize and leak
out the quantum dot. Consequently, electron-electron Coulomb
interactions becomes weak. As mentioned, $U_e$ and $U_h$ denote
the charging energies of QD for electrons and holes, respectively.
Therefore, the constant interaction model used in the Anderson
model is valid only for small electric field case, otherwise we
should take into account bias-dependent Coulomb interactions

\section{Formalism}
An Anderson model with two energy levels and constant interactions
is used to describe the system as shown in Fig. 1,

\begin{eqnarray}
H&=&\sum_{k} \epsilon_{k,e} a^{\dagger}_{k}a_{k}+\sum_{k}
\epsilon_{k,c} b^{\dagger}_{k}b_{k}+\sum_{k} \epsilon_{k,b}
c^{\dagger}_{k}c_{k}\\ \nonumber &+&\sum_{i=1,2} \epsilon_i
d^{\dagger}_{i} d_{i}+\lambda_{12} d^{\dagger}_{1} d_{2}+
\lambda^{*}_{21} d^{\dagger}_{2} d_{1}\\ \nonumber& +&\sum_{k,1}
t_{k,2}a^{\dagger}_{k}d_{2}+
+\sum_{k}t^{\dagger}_{k}b^{\dagger}_{k}d_{2}
+\sum_{k}t^{\dagger}_{k,1}c^{\dagger}_{k}d_{1}+h.c
\end{eqnarray}
where $a^{\dagger}_{k} (a_{k})$, $b^{\dagger}_{k} (b_{k})$ and
$c^{\dagger}_{k} (c_{k})$ create (destroy) an electron of momentum
$k$ in the emitter, collector and base electrodes, respectively.
The free electron model is considered in the electrodes in which
electrons have frequency-dependent energies
$\epsilon_{k,e(c)}=\varepsilon_k-\omega/2$ and
$\epsilon_{k,b}=\varepsilon_k+\omega/2$+v(t). Time-dependent
modulation v(t) denotes the time-dependent applied voltage in the
base electrode. $d^{\dagger}_{i}$ ($d_{i}$) creates (destroys) an
electron inside the QD with orbital energy
$\epsilon_i=E_i-(-1)^i\omega/2$. In this study $i=1$ and $i=2$
represent, respectively, the ground states of valence band and
conduction band of individual QDs. The fifth term describes the
coupling of the QD with electromagnetic field of frequency
$\omega$. $\lambda=-\mu_r {\cal E}$ is the Rabi frequency, where
$\mu_r=<f|r|i>$ is the matrix element for the optical transition
and ${\cal E}$ is the electric field per photon. $t_{k,i}$
describes the coupling between the band states of electrodes and
energy levels of QD. Note that a unitary transformation,
$S(t)=exp^{i\omega t/2(\sum_k
(c^{\dagger}_{k}c_{k}-a^{\dagger}_{k}a_{k}-b^{\dagger}_{k}b_{k})+d^{\dagger}_{1}
d_{1}-d^{\dagger}_{2} d_{2})}$, has been used to obtain Eq. (1)
via
\[
H=S^{-1}H(t)S-iS^{-1}\frac{\partial}{\partial t}S.
\]

To investigate the exciton assistant process, the interlevel
Coulomb interaction $U_{12}$ ($U_{eh}$) is taken into account in
Eq. (1)
\begin{equation}
H_U=U_{12} d^{\dagger}_1 d_1 d^{\dagger}_2 d_2,
\end{equation}
which is invariant under unitary transformation. Because we
restrict in the regime of applied voltage not sufficient to
overcome the charging energies resulting from $U_{ee}$ and
$U_{hh}$, therefore, we ignore $U_{ee}$ and $U_{hh}$ terms in this
study.$^{3}$

The emitter current can be calculated using the nonequilibrium
Keldysh Green's functions, which can be found in refs.[7,8]. Wang
et al have pointed out that the displacement current arising from
ac applied voltage is crucial to maintain gauge invariance, which
will be satisfied when the condition of total charge conservation
is satisfied.$^{9}$ The time-dependent emitter current is
consisted of the collector current and base current
$J_e(t)=J_c(t)+J_b(t)$.The collector current satisfying the charge
conservation and gauge invariance is given by
\begin{eqnarray}
J_c(t)&=&\frac{e}{\hbar}\frac{\Gamma_e
\Gamma_c}{\Gamma_e+\Gamma_c} \int \frac{d\epsilon}{\pi}
[f_e(\epsilon)-f_c(\epsilon)]\\ \nonumber &\times&
[-ImA^r_e(\epsilon,t)-\frac{1}{2}\frac{d|A^r_e(\epsilon,t)|^2}{dt}],
\end{eqnarray}
where $A^{r(a)}_e(\epsilon,t)=\int dt_1 e^{\pm
i\epsilon(t-t_1)}G^{r(a)}_e(t,t_1)$. $G^{r(a)}_e(t,t_1)$ denotes
the retarded (advanced) Green's function of electrons.  In Eq. (3)
we assume that tunnelling rates $\Gamma_{e(c)}=\sum_{k}
\delta(\epsilon-\epsilon_k)$ are energy-and bias independent. To
solve the spectral function of electrons $A^{r}_{e}(\epsilon,t)$,
the retarded Green's function of electrons $G^{r}_{e}(t,t_1)$ is
derived to obtain
\begin{eqnarray}
G^{r}_{e}(t,t_1)&=&(1-N_h(t_1))g^{r}_e(E_e,t,t_1)\\
\nonumber&+& N_h(t_1) g^{r}_e(E_e-U_{eh},t,t_1)
\end{eqnarray}
with
\begin{equation}
g^r_e(E_e,t,t_1)=-i\theta(t-t_1)e^{-i(E_e-i\Gamma_e/2)(t-t_1)},
\end{equation}
Two branches exist in Eq. (4) ; one corresponds to the electron
resonant energy level of $E_e$ with a weight of $(1-N_h(t_1))$,
and the other corresponds to the exciton resonant level of
$E_{ex}=E_e-U_{eh}$ with a weight of $N_h(t_1)$. Consequently,
electrons injected into the energy levels of QDs depends on not
only the emitter and collect voltages, but also on the hole
occupation number $N_h(t)$, which is given by
\begin{equation}
N_h(t)=\int\frac{d\epsilon}{\pi} \Gamma_h |A^r_h(\epsilon,t)|^2
\end{equation}
where $A^{r(a)}_h(\epsilon,t)=\int dt_1 e^{\pm
i\epsilon(t-t_1)}G^{r(a)}_h(t,t_1)$. The retarded Green's function
of holes is given by

\begin{equation}
G^r_h(E_h,t,t_1)=-i\theta(t-t_1)e^{-i(E_h-i\Gamma_h/2)(t-t_1)-i\int^t_{t_1}dt_2v(t_2)},
\end{equation}
where the time-dependent applied voltage v(t) denotes a
rectangular pulse with duration time $\Delta s$ and amplitude
$\Delta$.The time translation symmetry of $G^{r}_h(E_h,t,t_1)$ is
destroyed by v(t) (note that it is a typical n-i-n SET case).
Comparing to $G^{r}_h(E_h,t,t_1)$, the time translation symmetry
of $G^{r}_e(t,t_1)$ is destroyed by hole occupation number.
Therefore, it is expected that $I_c(t)$ will be very different
from the tunnelling current of typical n-i-n SETs.$^{10-12}$ The
expression of base current arising from the electron-hole
recombination of exciton state is given by
\begin{eqnarray}
& &J_{b}(t)\\ \nonumber &=&e \alpha \int d\omega~\omega^3 \int
\frac{d\varepsilon}{\pi^2}f^{<}_e(\varepsilon-\omega/2)
|A^a_{e}(\varepsilon-\omega/2,t)|^2 \\ \nonumber &\times&\Gamma_h
f_h(\varepsilon-\omega/2) |A^r_{h}(\varepsilon-\omega/2,t)|^2
\end{eqnarray}
where$f^{<}_e(\epsilon)=(\Gamma_e f_e(\epsilon)+\Gamma_c
f_c(\epsilon))$, $\alpha=4n^3_r \mu^2_r/(6c^3\hbar^3\epsilon_0)$,
where $n_r$ and $\epsilon_0$ are the refractive index and static
dielectric constant of system, respectively. We see that
$J_{b}(t)$ is determined by the time-dependent interband joint
density of states and the factors of $f_e(\epsilon)
f_h(\epsilon)$. To simplify the calculation of Eq. (8), we
approximate it as
\begin{equation}
J_b(t)=eR_{eh} N_e(t) N_h(t),
\end{equation}
where
\begin{equation}
N_e(t)=\int \frac{d\epsilon}{\pi} f^{<}_e(\epsilon)
|A^r_e(\epsilon,t)|^2.
\end{equation}
In Eq. (9) we define the time-independent spontaneous emission
rate $R_{eh}=\alpha \Omega^3_{ex}$, where
$\Omega_{ex}=E_g+E_e+E_h-U_{eh}$.

\section{Results and discussion}
Although tunnelling currents are more interesting than electron
and hole occupation numbers from the experimental viewpoint, we
still numerically  solve Eqs. (6) and (10) and show the occupation
number of electrons and holes as a function of time for different
amplitudes of applied rectangular pulse voltage with $\Delta
s=3t_0$at zero temperature in Fig. 4; the solid and dashed lines
denote, respectively, the amplitude $\Delta=30$ mV and $\Delta=20$
mV. When holes are injected into the QD, the exciton resonant
energy level for electrons is yielded. Consequently, the emitter
supplies electrons into the QD via the exciton resonant energy
level. In particular, some interesting oscillations superimpose on
the charing and discharging processes of holes. Such a oscillatory
behavior can not be observed for electron tunnelling process. In
addition, $N_e(t)$ exhibits a retarded response with respect to
time. For $t=1~t_0$, $N_e$ is still less than 0.2 although $N_h$
already reaches 0.8. This is because $N_e$ is in proportion to
$N^2_h(t)$ and the electron-channel behaviors as an opened system
since $\Gamma_e=\Gamma_c=0.5$~meV. On the other hand the hole
channel behaviors as a closed system since $R_{eh}<< \Gamma_h$.

Once electrons are injected into the QDs, the collector current
and base current occur.  Fig. 5(a) and (b) show the base current
and collector current, respectively. When electrons tunnel into
the QD from the emitter, photons are emitted from electron-hole
recombination of exciton state, such photons should exhibit the
antibounching feature with respect to time. Although in ref.[13]
the antibuching feature of photons was reported, electrons and
holes are injected into a single layer with dilute density of QDs.
Therefore, in ref.[13] the tunnelling current arising from the
spontaneous radiation of interband transition should be included
the particle size distribution.$^{14}$ It is worth noting that the
photon number correlation function used to examine the
antibunching characteristics are relevant with the base current
behavior.  Comparing to the base current, the exponential growth
and decay of collector current are not so faster as that of base
current. However, the collector current still mimics the behavior
of base current. Due to the collector current in the units of $ e
\times meV/\hbar$, the current gain defined as $\beta=J_c/J_b$ can
readily reach 100 for $\Gamma_{e}=\Gamma_c=0.5$meV. Consequently,
the response of small signal can be amplified through the output
of collector electrode. As mentioned, LESETs can be used as
efficient SPS detectors.

Finally, Figs. 6(a) and (b) show the base current and collector
current for two different tunnelling rates of hole. Other
parameters used are the same as those employed in Figs. (4) and
(5). Due to smaller tunnelling rate for solid line, hole
occupation number is smaller in solid line than that of dashed
line before pulse turns off. Subsequently, the electron occupation
number also becomes smaller for $\Gamma_h=0.5$ meV. Consequently,
the base current is suppressed for $t \le \Delta s=3 t_0$.
However, the discharging time is enhanced for electrons and holes
as $\Gamma_h=0.5$ meV, the base current at $t \ge \Delta s=3 t_0$
is larger at $\Gamma_h=0.5$ meV than $\Gamma_h=1$ meV. The
physical picture of collector current can be understood through
the interpretation of the base current. The results shown in Figs.
5 and 6 indicate that the shape of collector current can be tuned
by the combination of tunnelling rate and duration time of applied
signal. This could be used to manipulate the shape of
time-dependent current.

\section{Summary}
We have studied the dynamic properties of LESETs made from a
single QD embedded in a matrix connected with the emitter,
collector and base electrodes. Electrons are transport carriers in
the conduction band electrodes. As for valence band base
electrode, holes created by light excition or carrier doping play
a role of switch trigger for the base current and collector
current. The high current gain $\beta=J_c/J_b$ can be readily
reached using semiconductor engineering fabrication technique.
Such two outputs transistors exist potential application for the
next generation optoelectronics.$^{15}$

{\bf ACKNOWLEDGMENTS}

This work was supported by National Science Council of Republic of
China Contract Nos. NSC 94-2215-E-008-027 and NSC
94-2120-M-008-002.


\mbox{}

\newpage

{\bf Figure Captions}

Fig. 1. The schematic band diagram for the single quantum dot
(InAs) embedded in GaAs matrix connected with emitter, collector
and base electrodes. The exciton energy level $E_{ex}=E_e-U_{eh}$
is 35 meV above the Fermi energy of the emitter and collector
electrodes $E_{F,e}=50$ meV. The resonant energy level of holes is
15 meV above the Fermi energy of base electrode $E_{F,h}=50$ meV.
The voltage difference between the emitter and collector is
$V_{ec}=45$ mV. A small ac signal v(t) supplies holes into the
quantum dot.

Fig. 2: $U_{e}$, $U_{eh}$, and $U_{h}$ as functions of QD size for
biexciton (solid), negative trion $X^{-}$ (dotted) and positive
trion $X^{+}$ (dashed). Note that the ratio $(h-15 \AA)/(R_0-60
\AA)=1$ is used.

Fig. 3: Particle Coulomb interactions as functions of strength and
direction of electric field for QD with radius of $7.5$ nm and
height of $3$ nm. The solid line denotes the biexciton
configuration, and dashed and dotted lines, respectively, denote a
positive trion and a negative trion.

Fig. 4. Carrier occupation number as functions of time for two
different amplitude of a rectangular pulse with duration time
$\Delta s=3~t_0$. Time is in units of $t_0=\hbar/meV$. Temperature
$k_B T=0$ is considered throughout this study for simplicity.

Fig. 5. Diagram (a) and (b) correspond, respectively, for the base
current and collector current. All parameters used are the same as
those in Fig. 4. Base current and collector current are given in
units of $J_0=2e~R_{eh}$ and $e \times meV/\hbar$, where $R_{eh}$
denotes the spontaneous emission rate.

Fig. 6. Diagram (a) and (b) correspond, respectively, for the base
current and collector current. Solid line ($\Gamma_h=0.5$~meV) and
dashed line ($\Gamma_h=1.0$~meV). $\Delta=30$mV. Other parameters
used are the same as those in Fig. 5. Base current and collector
current are given in units of $J_0=2e~R_{eh}$ and $e \times
meV/\hbar$, where $R_{eh}$ denotes the spontaneous emission rate.


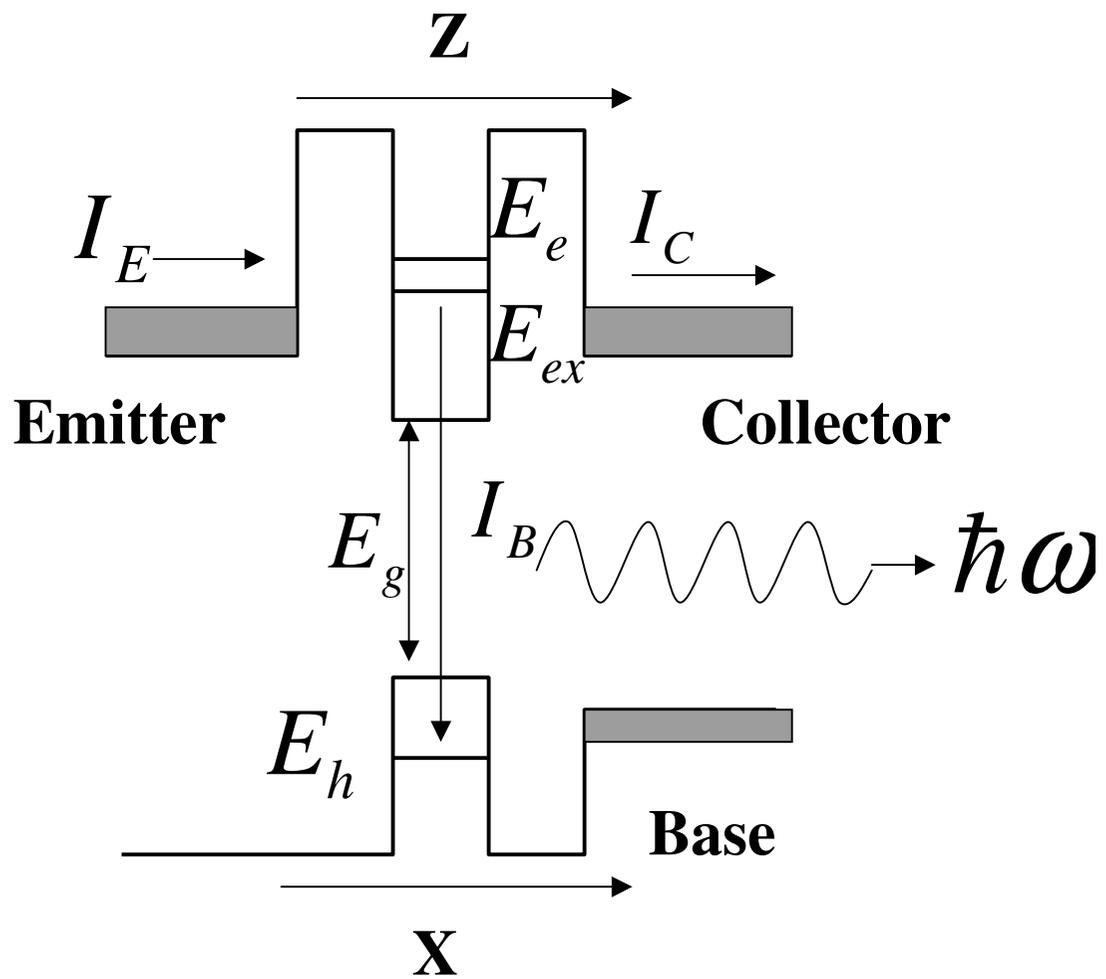
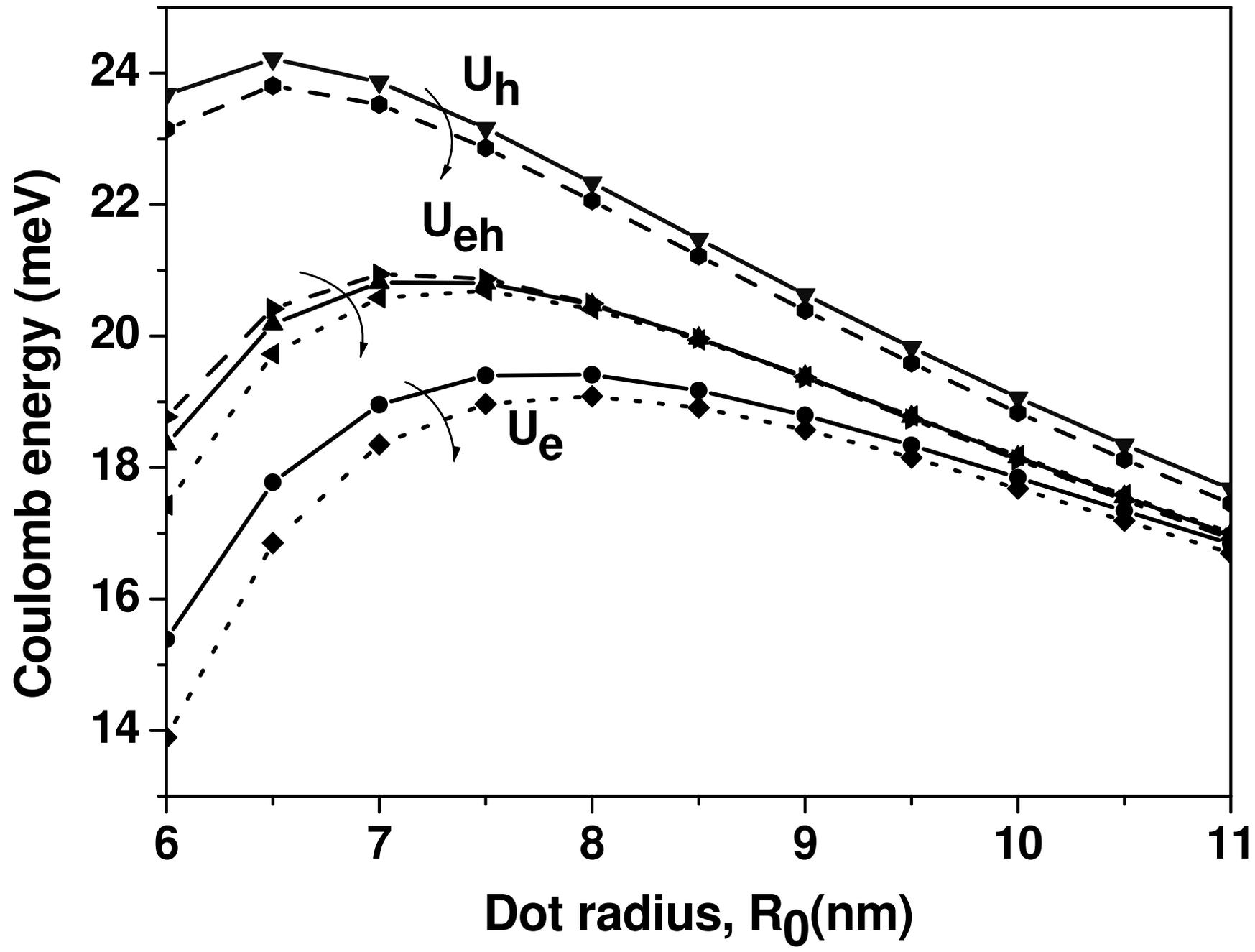
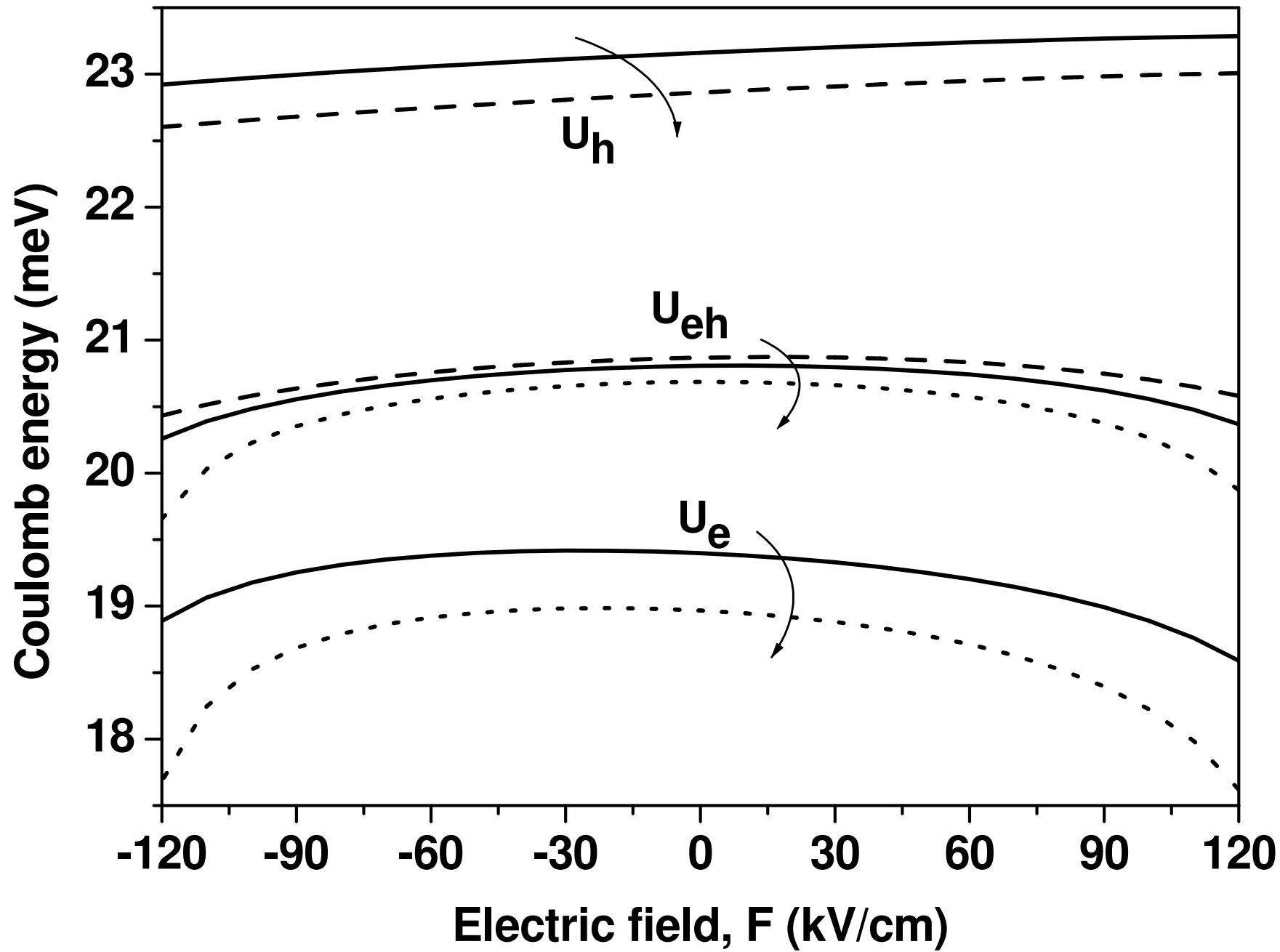
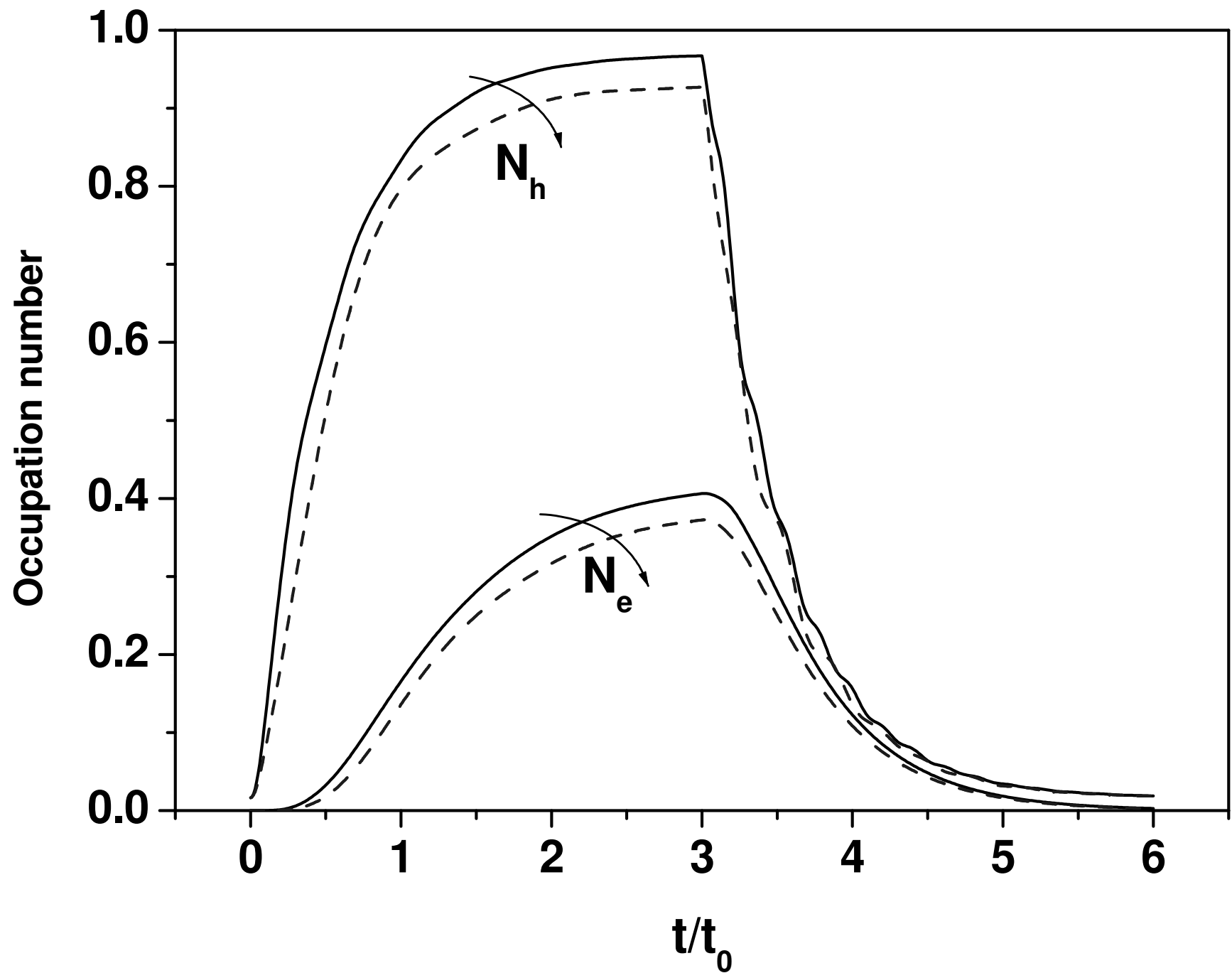
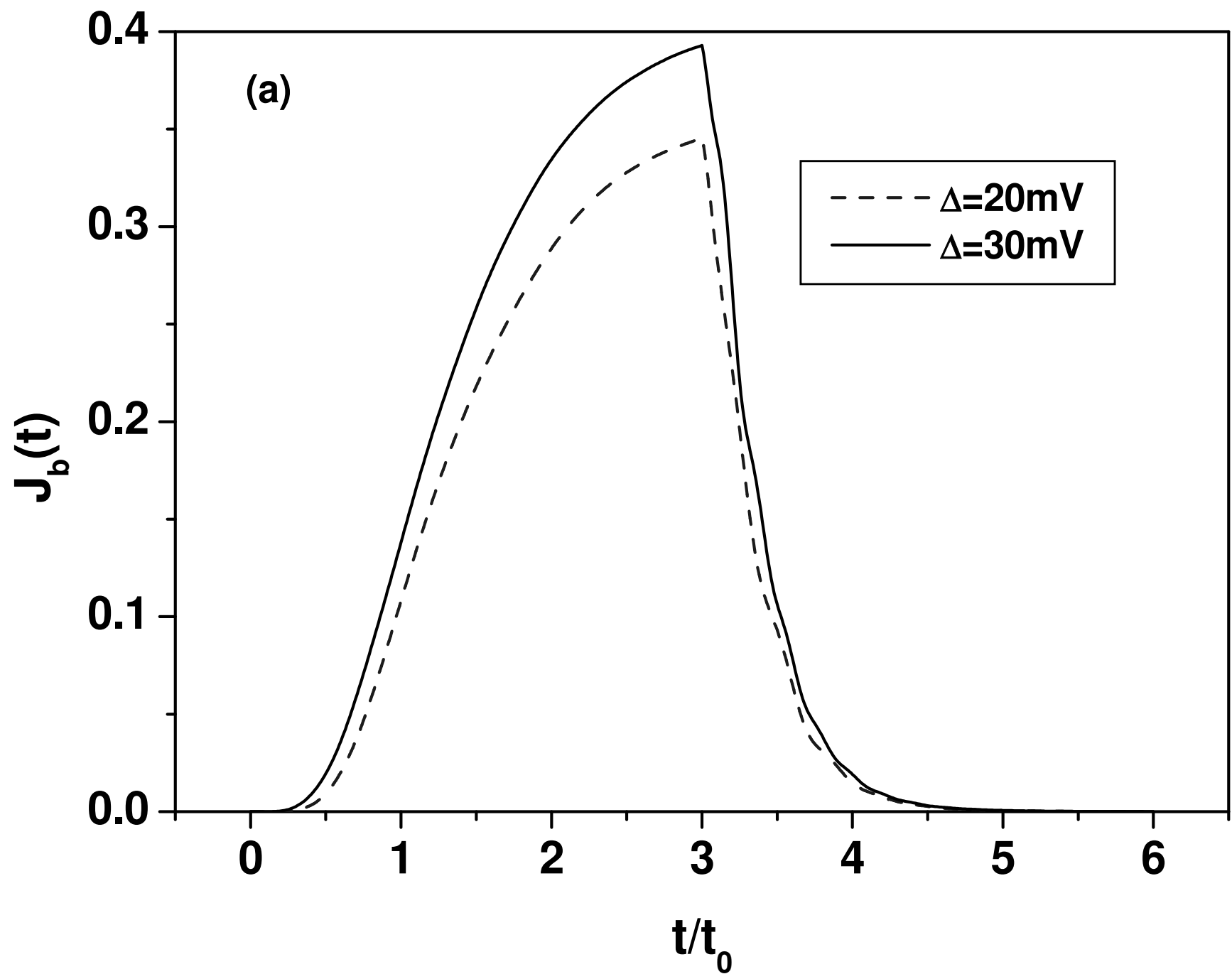
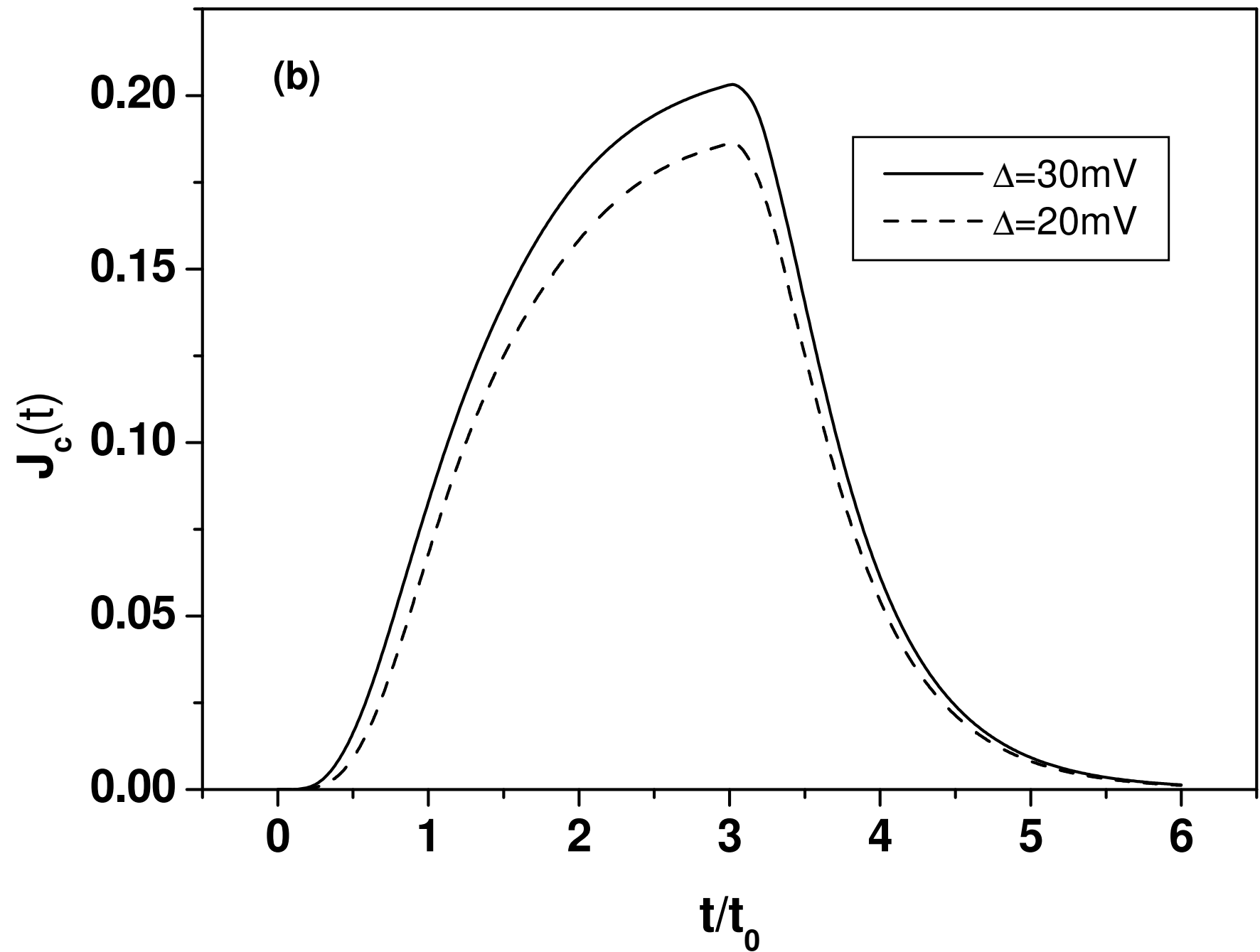
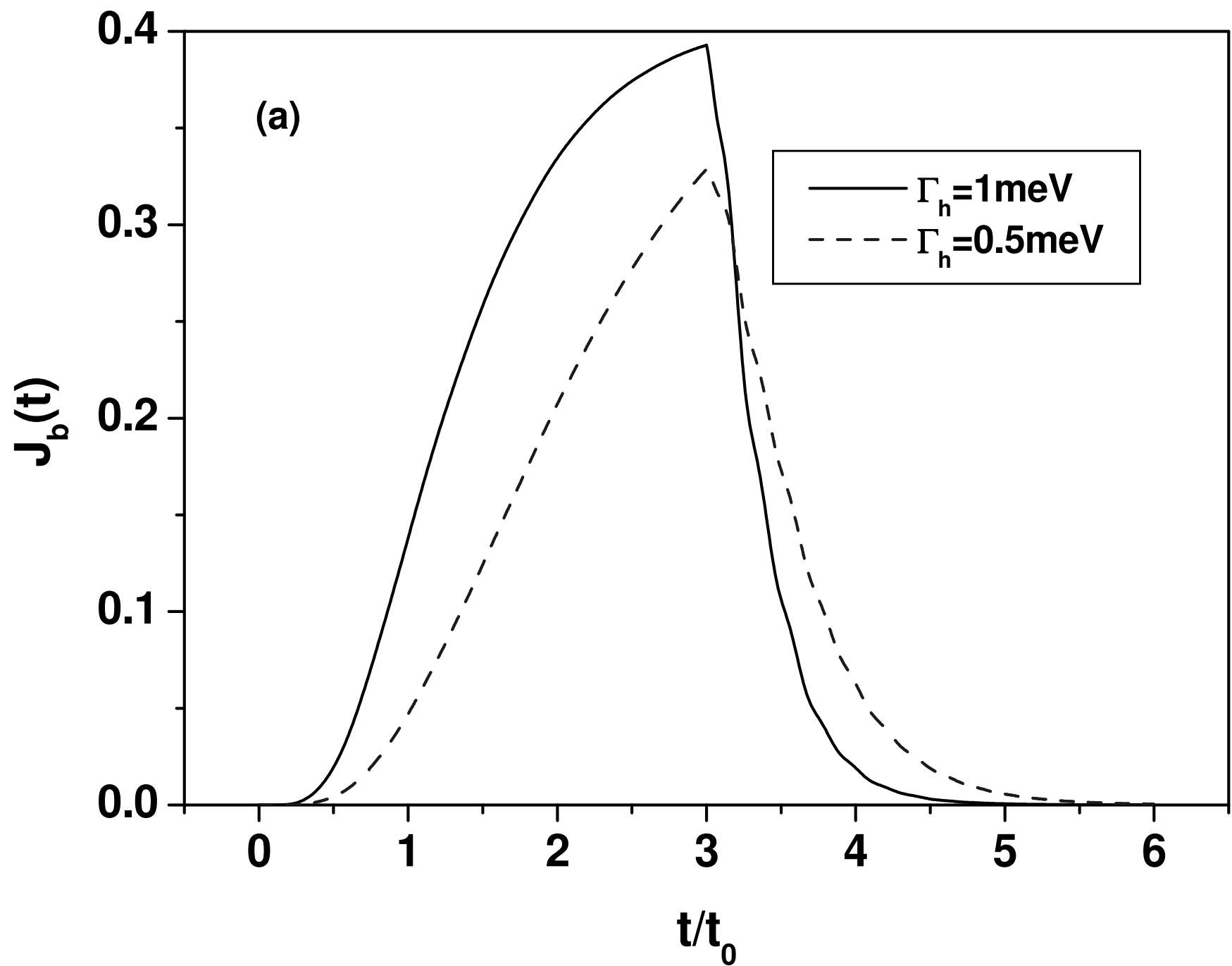
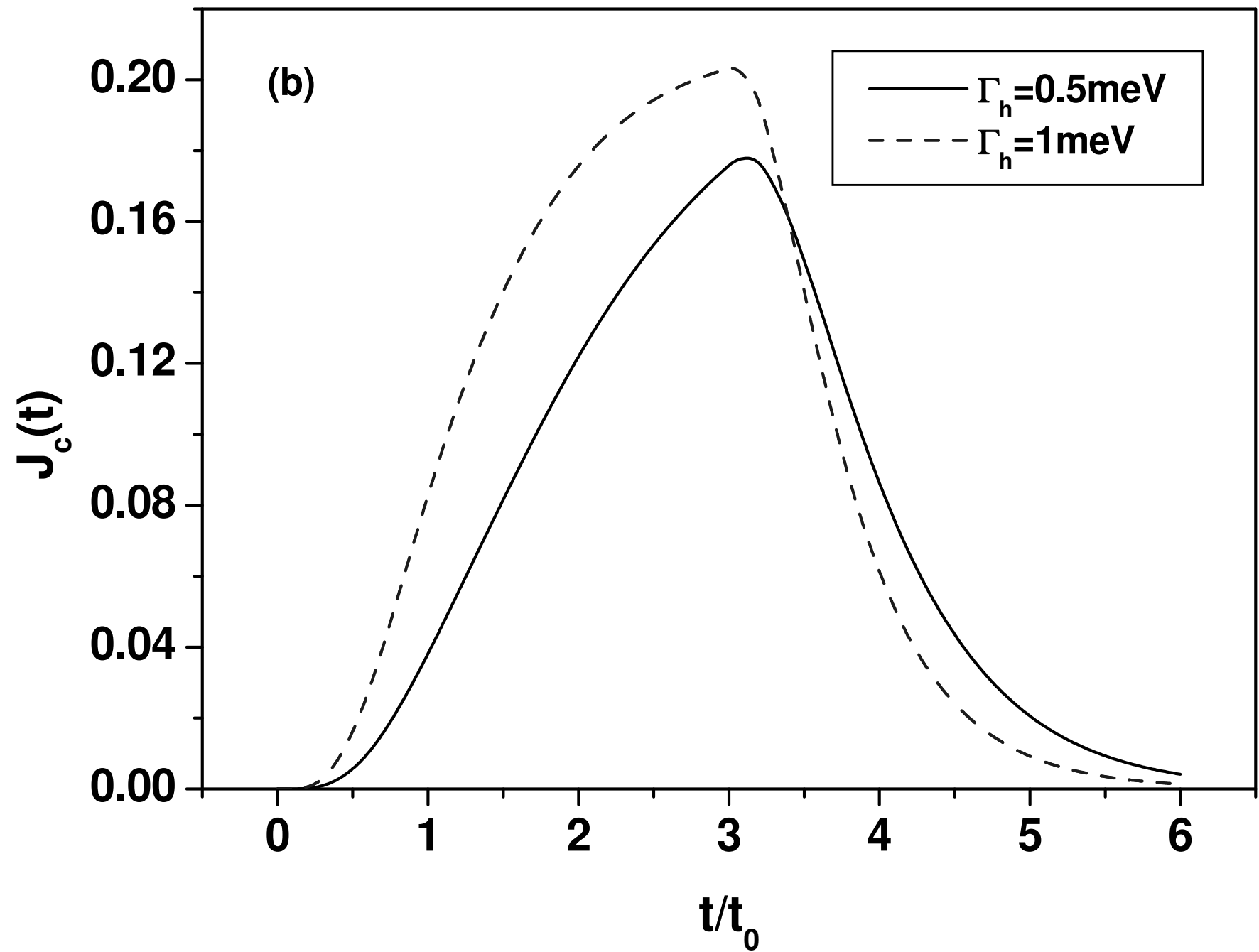
\end{document}